\documentstyle[12pt,epsfig]{article}

\newcommand{\lsim}{\raisebox{-0.13cm}{~\shortstack{$<$ \\[-0.07cm] $\sim$}}~}
\newcommand{\gsim}{\raisebox{-0.13cm}{~\shortstack{$>$ \\[-0.07cm] $\sim$}}~}

\newcommand{\tb}{\tan \beta}
\newcommand{\s}{\smallskip}
\newcommand{\nn}{\noindent}

\newcommand{\beq}{\begin{eqnarray}}
\newcommand{\eeq}{\end{eqnarray}}

\catcode`@=11
\def\citer{\@ifnextchar
[{\@tempswatrue\@citexr}{\@tempswafalse\@citexr[]}}
\def\@citexr[#1]#2{\if@filesw\immediate\write\@auxout{\string\citation{#2}}\fi
  \def\@citea{}\@cite{\@for\@citeb:=#2\do
    {\@citea\def\@citea{--\penalty\@m}\@ifundefined
       {b@\@citeb}{{\bf ?}\@warning
       {Citation `\@citeb' on page \thepage \space undefined}}%
\hbox{\csname b@\@citeb\endcsname}}}{#1}}
\catcode`@=12

\begin{document}

\vspace*{.1cm} 
\baselineskip=15pt

\begin{flushright}
CERN TH/2003--152\\
PM/03--15\\
July 2003\\
\end{flushright}

\vspace*{0.9cm}

\begin{center}

{\large\sc {\bf Detection of the neutral MSSM Higgs bosons}}

\vspace{0.3cm}

{\large\sc {\bf in the intense-coupling regime at the LHC}}

\vspace*{.7cm}

{\sc E. BOOS}$^{1}$, {\sc  A. DJOUADI}$^{2,3}$ and {\sc A. NIKITENKO}$^4$ 

\vspace*{5mm}

$^1$ 
Skobeltsyn Institute of Nuclear Physics, Moscow State University, \\
119992 Moscow, Russia.
 \vspace*{2mm} 

$^2$ Theory  Division, CERN, CH--1211 Geneva 23, Switzerland.
\vspace*{2mm}

$^3$ Laboratoire de Physique Math\'ematique et Th\'eorique, UMR5825--CNRS,\\
Universit\'e de Montpellier II, F--34095 Montpellier Cedex 5, France. 
\vspace*{2mm} 

$^4$ Imperial College, University of London, London, United Kingdom.
\end{center} 

\vspace*{1cm} 

\begin{abstract}

\nn We analyse the prospects to detect at the LHC the  neutral Higgs particles
of the Minimal Supersymmetric  Standard Model, when the masses of  the two
CP-even $h,H$ and of the CP-odd $A$ boson are close to one another, and the
value of  $\tb$ is large. In this ``intense-coupling  regime", the Higgs bosons
have strongly enhanced couplings to isospin down-type fermions and large total
decay widths, so that the $\gamma \gamma, WW^*$ and $ZZ^*$ decay modes of the 
three Higgs bosons are strongly suppressed. We advocate the use of the decays
into muon pairs, $h,H,A \to \mu^+ \mu^-$, to resolve the three Higgs boson
peaks: although the branching ratios are small, ${\cal O}(10^{4})$, the
resolution on muons is good enough to allow for their detection, if the mass
splitting is large enough. Using an event generator analysis and a fast
detector simulation, we show that only the process $pp \to b\bar{b} \mu^+
\mu^-$, when at least one of the $b$-quarks is detected, is viable.

\end{abstract}

\newpage 

\subsection*{1. Introduction}

The search for the Higgs bosons and the study of their fundamental properties
are the primary goals of the LHC. To make sure that Higgs particles with masses
in the vicinity of the electroweak scale cannot escape detection, two benchmark
models have been studied in great detail\,: the Standard  Model (SM), which
predicts the existence of a single Higgs particle $H^0$, and  its minimal
supersymmetric extension (MSSM), where the Higgs sector is extended to contain
two CP-even Higgs particles $h$ and $H$, a CP-odd or pseudoscalar Higgs boson
$A$, and two charged Higgs particles $H^\pm$ \cite{HHG}. In the case of the SM
Higgs particle, a plethora of production channels can be used at the LHC and
one of the main detection modes is expected to be the gluon--gluon fusion process,
$gg \to H^0$, with the signatures $H^0 \to \gamma \gamma$ or $H^0 \to ZZ^{(*)},
WW \to 4\ell$ in, respectively, the low and high Higgs boson mass ranges
\cite{LHCsearch}. \s

Because of the complexity of the Higgs spectrum, the situation is more 
complicated in the MSSM and depends on the values of the two input parameters
that characterize the Higgs sector at the tree level, the pseudoscalar Higgs
mass $M_A$ and the ratio of the vacuum expectation values of the two Higgs
doublet fields $\tb$. It depends also on the mixing in the scalar top sector,
which is controlled by the trilinear coupling $A_t$, when radiative 
corrections are included \cite{RCHiggs}. The latter push the maximal value of
the lightest $h$ boson from $M_{h}^{\rm max} \sim |\cos 2\beta|M_Z \leq M_Z$ at
the tree level, to $M_{h}^{\rm max} \sim 110$--130 GeV, depending on the 
values\footnote{Values $\tb \gsim 3$--10, depending on the mixing in the scalar
top sector, are required to maximize the $h$ boson mass and to evade the
experimental constraint from LEP2 searches, $M_h \simeq M_A \gsim 92$ GeV
\citer{LEPH}. From the theoretical  viewpoint, very large values of $\tb \sim
m_t/m_b \sim {\cal O} (50)$ are very interesting,  since they allow for Yukawa
coupling unification at the Grand Unification scale; see Ref.~\cite{Yukawa}.}
of $\tb$ and $A_t$.   The MSSM Higgs sector can be divided into three regimes,
according to the relative magnitudes of the pseudoscalar Higgs mass $M_A$ and 
the maximal value of the lightest $h$ boson mass $M_h^{\rm max}$. For large
$\tb$ values, for which $M_H^{\rm min} \simeq M_h^{\rm max}$, the search at the
LHC can be summarized as follows [for details, see Ref.~\cite{LHCsearch} for
instance]:\s

$(i)$ $M_A \gg M_{h}^{\rm max}$: in this case, we are in the so-called
decoupling regime,  in which the $H,A$ and $H^\pm$ bosons are very heavy and 
almost degenerate in mass, $M_A \sim M_H \sim M_{H^\pm}$, while the $h$ boson 
has a mass $M_h \simeq M_h^{\rm max}$ and SM-like Higgs properties. The 
techniques devised for the detection of a light $H^0$ particle can be adapted
to the $h$ boson. The $A$ and $H$ bosons, if not too heavy, can be searched for
in the  channels $gg/q \bar q \to b\bar{b}H/A \to b\bar{b} \tau^+ \tau^-$,
while the $H^\pm$ particle can be detected in the process $gg/q\bar q  \to
t\bar{b}H^- \to t\bar{b} \tau \nu_\tau$ for instance.\s

$(ii)$ $M_A < M_h^{\rm max}$: in this case, it is the heavier $H$ boson that
will be SM-like, while the $h$ and $A$ particles will be degenerate in mass and
couple strongly to $b$-quarks and $\tau$-leptons for large values of $\tb$. The
search techniques are the same as above, except that the roles of the $h$ and
$H$ bosons are interchanged. The $H^+$ particles, because of the MSSM relation 
$M_{H^\pm} \sim \sqrt{M_A^2 +M_W^2}$, are light enough to be detected  in 
top-quark decays, $t \to H^+ b$.\s

$(iii)$ $M_A \sim M_{h}^{\rm max}$: this is what was called the  
``intense-coupling  regime" \cite{intense} where the three neutral Higgs
bosons  have comparable masses, $M_h \simeq M_H \simeq M_A$. The couplings of
the CP-even Higgs particles to gauge bosons are both suppressed with respect to
the SM [the $A$ boson does not couple to  gauge bosons because of
CP-invariance], while the $h,H$ and $A$ couplings to down-type (up-type)
fermions are strongly enhanced (suppressed). \s

While detailed experimental analyses have been performed for the first two
scenarios \cite{LHCsearch}, only little work has been done for the 
intense-coupling regime. In the detailed theoretical discussion given in
Ref.~\cite{intense}, it was shown that the search at the LHC might be rather
difficult in this regime. The main problem is due to the fact that, for
$M_h\sim M_H \sim M_A$ and $\tb \gg1$, the three neutral Higgs bosons will
mainly decay into isospin down-type fermions and the clear $\gamma \gamma$ and
$ZZ^*,WW$ signatures cannot be used anymore, the branching fractions being too
small. In addition, since the Higgs masses are close, it will be difficult to
detect individually the three Higgs bosons, and resolving between the  peaks is
made even more difficult since the total decay widths can be rather large,
implying broader signals. \s

In this note, we  discuss the detection of the three neutral MSSM Higgs bosons
$\Phi=h,H,A$ in this scenario, paying a special attention to the possibility of
resolving the signal peaks.  Performing an event generator analysis that takes
into account the signals and the various backgrounds, as well as  a simulation
of some aspects of one of the LHC detectors [CMS] response, we show that the
detection of separate Higgs bosons can be extremely  difficult. It can be done
only if the Higgs mass differences are sizeable and only if the rare decays
into muon pairs, $\Phi \to \mu^+ \mu^-$, which have branching ratios at the
level of a few times  $10^{-4}$, are exploited. In addition, the Higgs bosons
need to be produced in the  associate processes $gg \to b\bar{b} \Phi$, i.e.
with at least one $b$-quark being detected; the  processes $gg \to \Phi$ and
$b\bar{b} \to \Phi$ suffer from the very large background from Drell--Yan
production $pp\to \gamma^*,Z^{(*)}\to\mu^+ \mu^-$. \s

The rest of the discussion is as follows.  In the next section, we will recall
the main features of the intense coupling scenario. In section 3, we discuss
the production of the neutral Higgs bosons and their main backgrounds at the
LHC. In section 4, an event generator analysis for the separation of the Higgs
bosons is presented. A short conclusion is then given. 

\subsection*{2. The intense-coupling regime}

As introduced above, the intense-coupling regime is characterized by a rather
large value of $\tb$, and a pseudoscalar Higgs boson mass that is close to the
maximal (minimal) value  of the CP-even $h$ ($H$) boson mass, $M_A \sim
M_h^{\rm max}$, almost leading to a mass degeneracy of the neutral Higgs
particles, $M_h \sim M_A \sim M_H$. In the following, we will summarize the
main features of this scenario. For the numerical illustration, we will fix the
parameter $\tb$ to the value $\tb=30$ and choose the maximal mixing scenario,
where the trilinear Higgs--stop coupling is given by  $A_t \simeq \sqrt{6} M_S$
with the common stop masses fixed to $M_S=1$ TeV; the other SUSY parameter will
play only a minor role. The determination of the Higgs masses, couplings and
branching ratios is performed using  the program {\tt HDECAY} \cite{hdecay} in
which  the routine {\tt FeynHiggsFast} \cite{FHF} is used for the
implementation of the radiative corrections. \s

The left-hand side of Fig.~1 displays the masses of the MSSM Higgs bosons  as
a function of $M_A$, the latter varying from 100 to 140 GeV for our 
representative value of $\tb$ in the scenario of maximal stop mixing. As  can
be seen, for $M_A$ close to the maximal $h$ boson mass, which in this case is
$M_h^{\rm max}  \simeq 130$ GeV, the mass differences $M_A-M_h$ and $M_H - M_A$
are less than about 5 GeV. The $H^\pm$ boson mass, given by $M_{H^\pm}^2 \sim
M_A^2 +M_W^2$, is larger\,: in the range $M_A \lsim 140$ GeV, one has
$M_{H^\pm} \lsim 160$ GeV,  implying that charged Higgs bosons can always be
produced in top-quark decays, $t \to H^+ b$, and be detected at the LHC.\s

The couplings of the CP-even Higgs bosons to fermions and gauge bosons 
normalized to the SM Higgs boson couplings are shown in the right-hand side of
Fig.~1 for the same inputs as previously. For small $M_A$ values, the $H$ boson
has almost SM couplings, while the couplings of the $h$ boson to $W,Z,t$ $(b)$ 
are suppressed (enhanced); for large $M_A$ values the roles of $h$ and $H$ are
interchanged. For medium values, $M_A \sim M_h^{\rm max}$, the couplings of
both $h$ and $H$ to gauge bosons $V=W,Z$  and top quarks are suppressed, while
the couplings to $b$ quarks [for which $10 \times g_{\Phi bb}^{-2}$ are shown  
in the figure] are strongly enhanced. The normalized couplings of the CP-even
Higgs particle are simply $g_{AVV}=0$ and $g_{Abb} = 1/ g_{Att} =\tan\beta
=30$.\s

\begin{figure}[htbp]
\begin{center}
\vspace*{-1.7cm}
\centerline{\psfig{file=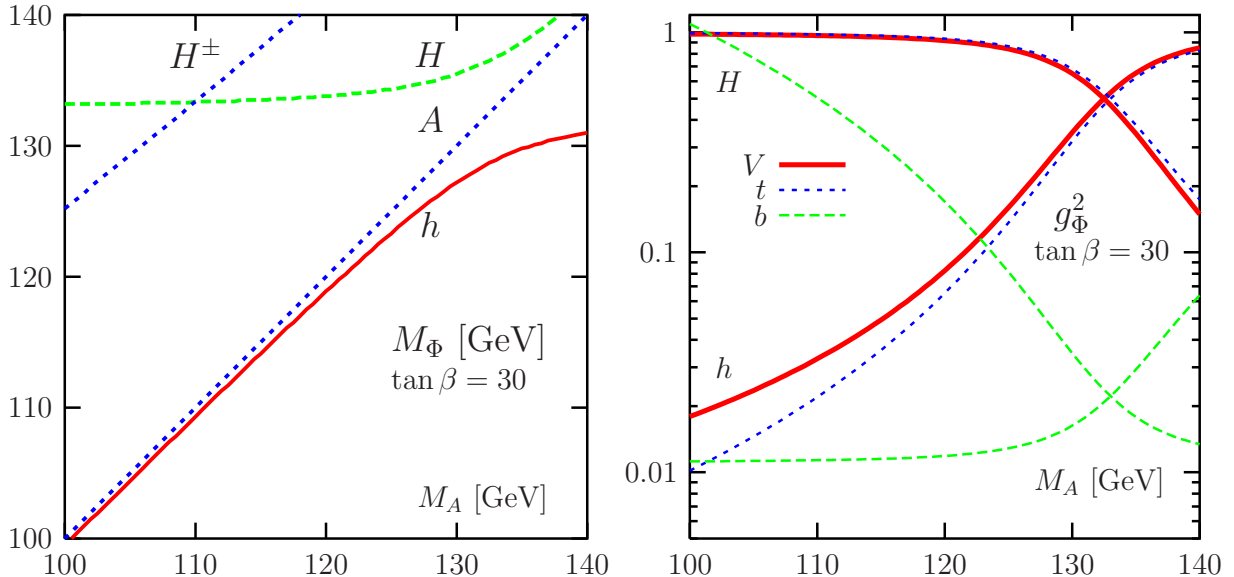,width=20cm}}
\vspace*{-19.7cm}
\caption{\it The masses of the MSSM Higgs bosons (left) and the normalized 
couplings of the CP-even Higgs bosons to vector bosons and third-generation 
quarks (right) as a function of $M_A$ and $tan \beta=30$. For the $b$-quark
couplings, the values $10 \times g_{\Phi bb}^{-2}$ are plotted.}
\end{center}
\vspace*{-1.cm}
\end{figure}

These couplings determine to a large extent the branching ratios of the Higgs
particle decays, which are shown in Fig.~2. Because the couplings of the three
Higgs particles to $b$-quarks and $\tau$-leptons are strongly enhanced, their
branching ratios to $b\bar{b}$ and $\tau^+\tau^-$  final are the dominant ones,
with values $\sim 90$\% and $\sim 10$\% respectively. The decays $H \to WW^*$
do not exceed the level of 10\%, even for small $M_A$ values [where $H$ is
almost SM-like] and in most of the range displayed for $M_A$, both the decays
$H,h \to WW^*$ are suppressed to the level where they are not useful.  The
decays into $ZZ^*$ are one order of magnitude smaller.  The interesting rare
decay mode into $\gamma \gamma$, which is at the level of  a few times
$10^{-3}$ in the SM, is very strongly suppressed for the three Higgs particles
and cannot be used anymore. Finally, note that the  branching ratios for the
decays into muons, $\Phi \to \mu^+ \mu^-$, are constant in the entire $M_A$
range exhibited, at the level of $3 \times 10^{-4}$. \s

\begin{figure}[htbp]
\hspace*{-1cm}
\begin{center}
\vspace*{-2.cm}
\centerline{\psfig{file=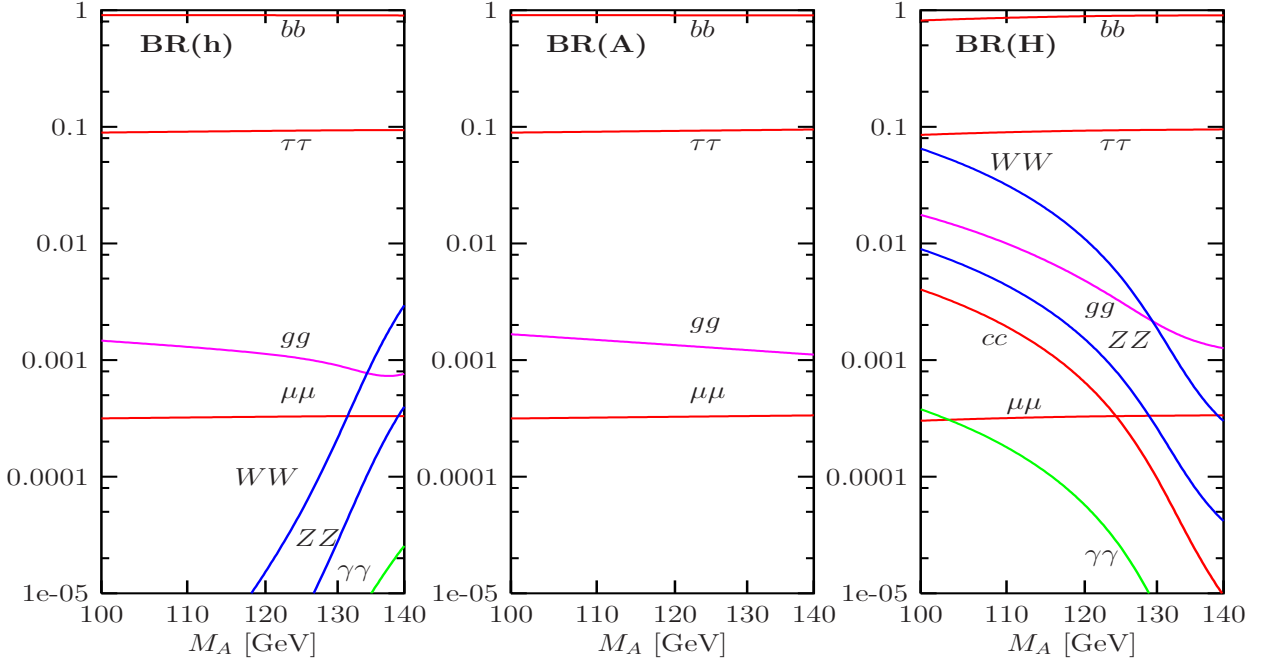,width=20cm,,height=23cm}}
\vspace*{-12.3cm}
\caption{\it The branching ratios of the neutral MSSM Higgs bosons $h,A,H$ for 
the various decay modes as a function of $M_A$ and for $tan \beta=30$. }
\end{center}
\vspace*{-1.cm}
\end{figure}

Summing up the partial widths for all decays, the total decay widths of the
three Higgs particles are shown in the left-hand side of Fig.~3. As can be
seen, for $M_A \sim 130$ GeV, they are at the level of 1--2 GeV, i.e. two
orders of magnitude larger than the width of the SM Higgs boson for this value
of $\tb$ [the total width increases as $\tan^2\beta$]. The right-hand side of
the figure shows the mass bands $M_\Phi \pm \Gamma_\Phi$ and, as can be seen, 
for the above value of $M_A$, the three Higgs boson masses are overlapping.\s

\begin{figure}[!ht]
\begin{center}
\vspace*{-1.5cm}
\centerline{\psfig{file=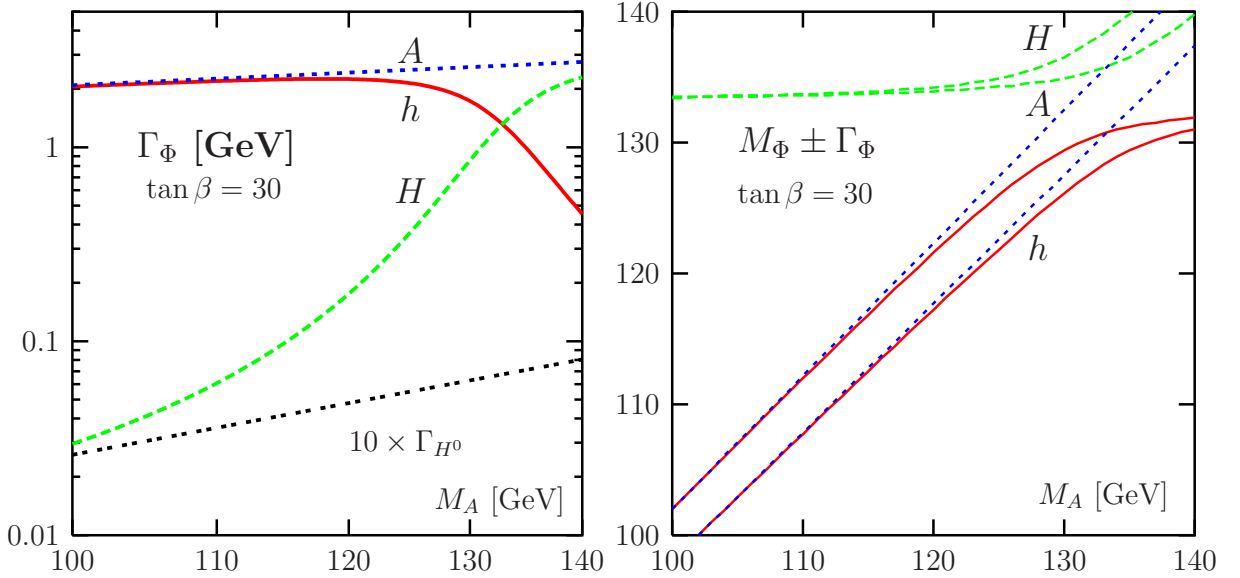,width=20cm}}
\vspace*{-19.7cm}
\caption{\it  Total decay widths $\Gamma_\Phi$ (left) and the mass bands $M_\Phi
\pm \Gamma_\Phi$ (right) for the neutral MSSM Higgs bosons as a function of 
$M_A$ and for $tan \beta=30$.}
\end{center}
\vspace*{-1.1cm}
\end{figure}

All these points are summarized for the three values $M_A =125, 130 $ and 135 
GeV in Table~1, where we display the Higgs boson masses, their total decay
widths and the branching ratios of some important decay modes. These three
points, called respectively P1, P2 and P3, are the ones we will choose to
perform our analysis of  the production and the detection of these Higgs
particles at the LHC, the subject to which we now turn our attention. \s

\begin{table}[htbp]
\renewcommand{\arraystretch}{1.2}
\begin{center}
\vspace*{-7mm}
\begin{tabular}{|c|c||c|c||c|c|c||} \hline
Point & $\Phi$ & $M_\Phi$  & $\Gamma_\Phi$ & BR$(\gamma \gamma)$ &
BR($WW^*$) & BR $(\mu^+ \mu^-)$\\ \hline
  &$h$ & 123.3 & 2.14 & $1.9 \times 10^{-6}$ & $5.2 \times 10^{-5}$ & 
  $3.29 \times 10^{-4}$ \\
P1&$A$ & 125.0 & 2.51 & $5.9 \times 10^{-7}$ & $0$ & $3.29 \times 10^{-4}$ 
\\
  &$H$ & 134.3 & 0.36 & $2.4 \times 10^{-5}$ & $5.1 \times 10^{-3}$ & 
  $3.31 \times 10^{-4}$ \\
\hline
  &$h$ & 127.2 & 1.73 & $3.7 \times 10^{-6}$ & $2.1 \times 10^{-4}
  $ & $3.30 \times 10^{-4}$ \\
P2&$A$ & 130.0 & 2.59 & $4.7 \times 10^{-7}$ & $0$ & $3.31 \times 10^{-4}$ \\
  &$H$ & 135.5 & 0.85 & $7.4 \times 10^{-6}$ & $1.9 \times 10^{-3} $ & 
  $3.33 \times 10^{-4}$ \\
\hline
  &$h$ & 129.8 & 0.97 & $1.0 \times 10^{-5}$ & $9.2 \times 10^{-4}
  $ & $3.31 \times 10^{-4}$\\  
P3&$A$ & 135.0 & 2.67 & $4.5 \times 10^{-7}$ & $0$ & $3.33 \times 10^{-4}$ 
\\
  &$H$ & 137.9 & 1.69 & $1.8 \times 10^{-6}$ & $6.5 \times 10^{-4}$ & 
  $3.35 \times 10^{-4}$ \\
\hline
\end{tabular}
\end{center}
\vspace*{-5mm}
\caption[]{\it Masses, total widths (in GeV) and some decay branching ratios 
for the points P1, P2 and P3. The cross sections for the processes $pp \to \Phi$
and $\Phi b\bar{b}$ (in pb) are also shown.}
\vspace*{-2mm}
\end{table}

\subsection*{3. Signals and backgrounds at the LHC}

As discussed above, the most difficult problem  we must face in the
intense-coupling regime, is to resolve between the three peaks of the neutral 
Higgs bosons when their masses are close to one another. The only decays with
large branching ratios on which we can rely are the $b\bar{b}$ and  $\tau^+
\tau^-$ modes. While the former has too large a QCD background to be useful,
the latter channel has been shown to be viable for  discovery\footnote{Note
that in previous CMS analyses of the $pp \to b\bar{b}+ \mu^+ \mu^-$ signature,
the complete 4-fermion background was not been fully included. This remark 
applies also to the $pp \to b\bar b + \tau^+ \tau^-$  signature.}
\cite{LHCsearch}. However, the expected experimental resolution on the
invariant mass of the  $\tau^+ \tau^-$ system, in the mass range that we are
interested in, is of the order of 10 to 20 GeV, and thus clearly too large to
resolve the three Higgs peaks. One would then simply observe a relatively wide
resonance corresponding to $A$ and $h$ and/or $H$ production.  Since the
branching ratios of the decays into $\gamma \gamma$ and $ZZ^* \to  4\ell$ are
too small, a way out is  to use the Higgs decays into muon pairs: although the
branchings ratio are rather small, BR($\Phi \to \mu^+\mu^-) \sim 3.3 \times
10^{-4}$,  the resolution is expected to be as good as 1 GeV, i.e. comparable
to the total width,  for $M_\Phi \sim 130$ GeV. \s

Because of the strong enhancement of the Higgs couplings to bottom quarks, the
three Higgs bosons  will be produced at the LHC mainly in the
gluon--gluon process
\beq
gg \to \Phi=h,H,A \to \mu^+ \mu^- \ , 
\eeq
which is dominantly mediated by $b$-quark loops, and the associated production 
with $b\bar{b}$ pairs, 
\beq
gg/q\bar{q} \to b\bar{b}+\Phi =h,H,A \to b\bar{b}+\mu^+ \mu^- \ .
\eeq
The  Higgs-strahlung $pp \to HV$ and vector-boson fusion  $qq \to H qq$
processes, as well as associated production with top quarks, will have smaller
cross sections than for the SM Higgs boson since the couplings to the involved
particles are suppressed. The two processes eqs.~(1) and (2) have recently been 
discussed in Refs.~\cite{Tao} and \cite{Sally}, respectively, and have been shown
to be viable for the discovery of relatively light Higgs bosons\footnote{The 
vector-boson fusion process $qq \to H^0 \to \mu^+ \mu^-$ has also been considered
for a SM  Higgs boson \cite{Tilman} and has been shown to work at the LHC only
if an unreasonable amount of luminosity is collected. In addition, the process 
$qq  \to h,H \to \tau^+ \tau^-$ discussed in Ref.~\cite{WWfusion} cannot be
used to resolve the two $h$ and $H$ peaks in the intense-coupling regime,
because of the poor resolution on the $\tau^+ \tau^-$ invariant mass.}. \s 

For the calculation of the signal cross sections for the $pp \to \mu^+ \mu^-$
final state, since they give rise to the same topology, we sum the cross
sections of both the $gg \to \Phi \to \mu^+ \mu^-$ process and the $gg/q\bar{q}
\to b\bar{b}+\Phi$ process where the transverse momenta of the $b$-quarks are
too small , $p_\perp^{b,\bar{b}} \leq 20$ GeV, to be detected. For the former
process, we apply a  $K$-factor of 1.7 \cite{K-fac} to take into account the
NLO QCD corrections, while for the latter one, we apply a $K$-factor of 1.5 to
the LO cross section evaluated at a scale $\mu \sim \frac{1}{2} M_H$   as
recently reported\footnote{Note that the cross section calculated  directly for
bottom-quark fusion $b\bar{b} \to \Phi$ at NNLO \cite{Robert} gives
approximately the same  results if the factorization and the renormalization 
scales are chosen properly,  as it was shown in  Refs.~\cite{Scott,Bdensity}.}
in Ref.~\cite{Michael}. One then obtains production rates at the level of a
fraction of a picobarn for $pp \to h,H,A \to \mu^+ \mu^-$.  For the process $gg
\to \Phi b\bar{b}$, where we require the two bottom quarks to  be detected,
i.e. with  $p_\perp^{b, \bar{b}} \geq 20$ GeV, the NLO cross section is
approximately the same as the  LO cross section when it is evaluated at the
scale $\mu=M_H/2$ and a running $b$-quark mass for the bottom Yukawa coupling
\cite{Michael}.   The obtained cross sections in this case are one order of
magnitude smaller than those of the inclusive $pp \to \Phi \to \mu^+ \mu^-$
case. \s 

For the backgrounds to $\mu^+\mu^-$ production, we have included only the
Drell--Yan process $pp \to  \gamma^*, Z^* \to \mu^+\mu^-$, which is expected 
to be the largest source [as will be seen later, this is sufficient for our
conclusion]. For the $pp  \to \mu^+ \mu^- b\bar{b}$ final state, however, we
have included the full 4-fermion background, which is mainly due to the process
$pp \to b\bar{b} Z$ with $Z \to \mu^+ \mu^-$. Both signals and backgrounds 
have been generated with the program {\tt CompHEP} \cite{comphep}. \s

\begin{figure}[!ht]
\begin{center}
\vspace*{-1.5cm}
\centerline{\psfig{file=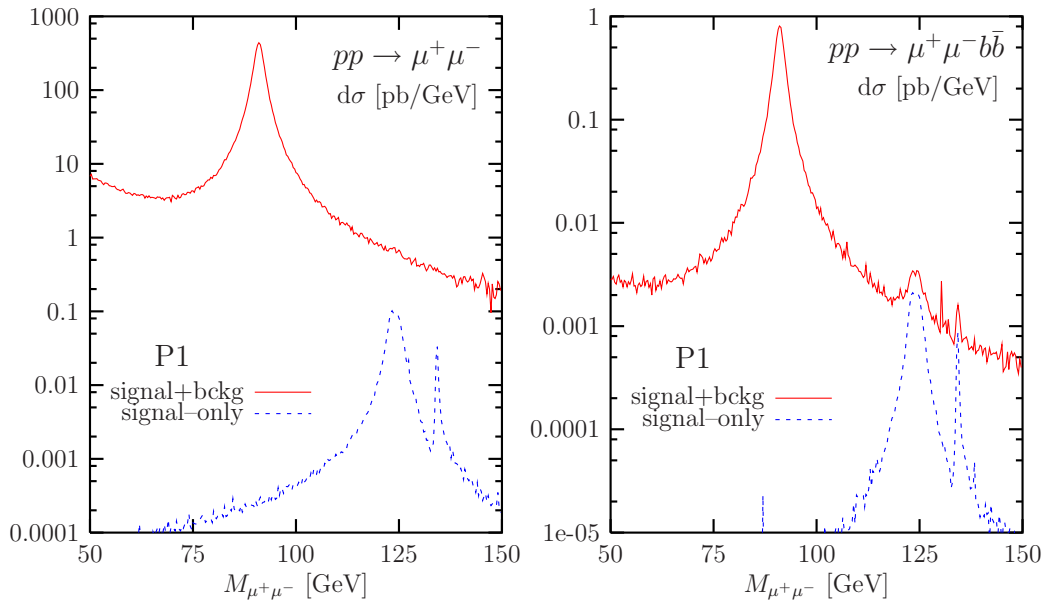,width=17cm}}
\vspace*{-15.5cm}
\caption{\it  The differential cross section in pb/GeV as a function of the
dimuon mass for the point P1, for both the signal and signal plus background 
in the processes $pp (\to \Phi) \to \mu^+ \mu^-$ (left figure) and $pp (\to \Phi 
b\bar{b}) \to \mu^+ \mu^- b\bar{b}$ (right figure).}
\end{center}
\vspace*{-1cm}
\end{figure}

\newpage

The differential cross sections are shown for the scenario P1 as a function of 
the invariant dimuon mass in the left-hand side of Fig.~4, for the final state
$pp (\to h,H,A) \to \mu^+ \mu^-$. As can be seen, the signal rate is fairly
large and we may hope, in principle, to see two peaks: one corresponding to the
production of the $h/A$ bosons and one to the production of the $H$ boson. 
However, when put on top of the huge Drell--Yan background, the signal becomes
completely invisible. This holds true even with some optimization of the cuts;
for instance, applying a cut on the muon transverse momenta $p_\perp^\mu \geq
50$ GeV, which should strongly suppress the Drell--Yan cross section, still
leaves too large a background.  The same features hold for the points P2 and
P3, and we conclude, contrary to Ref.~\cite{Tao}, that already at the level of
a ``theoretical simulation", the  Higgs boson signal in the  inclusive $pp \to
\mu^+ \mu^-$ process will probably be hopeless to extract for Higgs masses
below 140 GeV, unless  unreasonably large values of the parameter $\tb$ are
chosen to enhance the signal rate. \s

In the right-hand side of Fig.~4, we display, again for scenario P1, the signal
cross section from $pp \to \mu^+\mu^- b\bar{b}$ and the complete 4-fermion SM 
background cross section as a function of the dimuon system. The number of
signal events is an order of magnitude smaller than in the previous case, but
one can still see the two peaks. However, the main difference here is that the
rate for the  background is much smaller than in the Drell--Yan case. Once the
signal  events are put on top of the background distribution, one  can still
see the two peaks  corresponding to $h/A$ and $H$ production. The same analysis
has been repeated for scenarios P2 and P3, and the output for the signal and 
background rates is shown in Fig.~5. As can be seen, the situation is similar
to that of the previous case, except that here the mass difference between the
Higgs bosons is large enough for us to hope that the three individual peaks 
could be resolved. \s

Note, however, that up to now, we did not include any efficiency for the
detection of $b$ quarks and muons and did not assume any resolution for the
mass of the dimuon system.

\begin{figure}[!ht]
\begin{center}
\vspace*{-1.5cm}
\centerline{\psfig{file=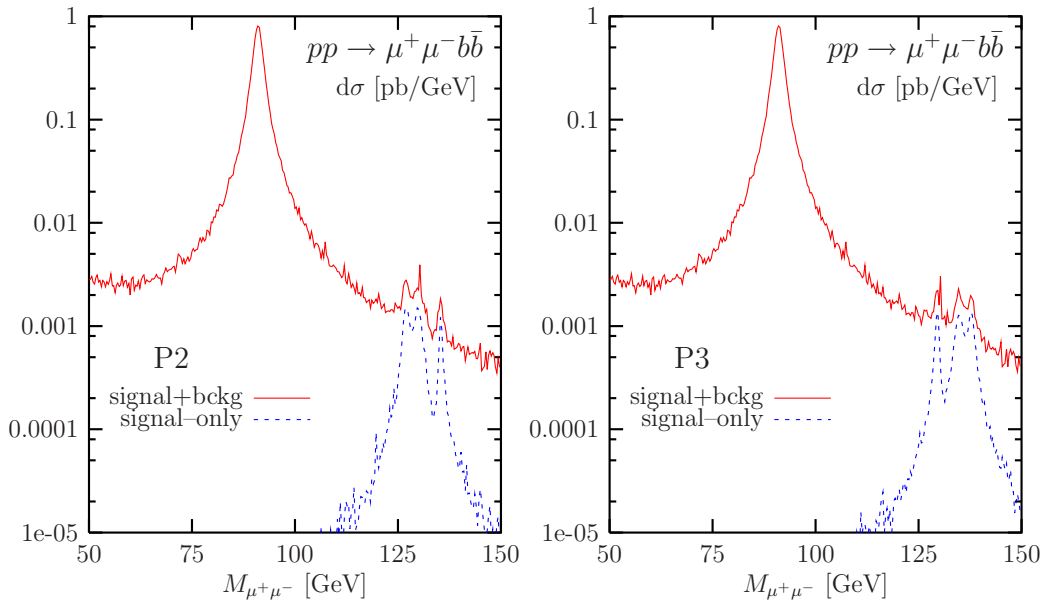,width=17cm}}
\vspace*{-15.5cm}
\caption{\it  The differential cross section in pb/GeV as a function of $M_A$ 
for  both the signal and signal plus background in the process $pp (\to \Phi 
b\bar{b}) \to \mu^+ \mu^- b\bar{b}$ for P2 (left) and P3 (right).}
\end{center}
\vspace*{-1.1cm}
\end{figure}
\newpage

\subsection*{4. An event generator analysis}

In order to perform a more realistic analysis, we have generated unweighted
events for the full 4-fermion background $pp \to \mu^+ \mu^- +b\bar{b}$ and for
the signal, for the three parameter points P1, P2 and P3 already introduced; as
above, the generator {\tt CompHEP} has been used. With the help of the new {\tt
CompHEP-PYTHIA} interface \cite{interface}, which has been upgraded to include
the implementation of the so-called Les Houches Accord 1 \cite{accord1}, the
unweighted events have been processed through {\tt PYTHIA 6.2} \cite{pythia}
for fragmentation and hadronization, taking into account initial- and
final-state radiation. We stress, once more, that the requirement of observing
both $b$-jets, with $p_\perp^{b,\bar{b}} >20$ GeV,  leads to a reduction by a
factor of 10 of the signal rate compared to the fully inclusive  case. In turn,
the background reduction factor in this case is about 200, resulting in a much
larger $S/\sqrt{S+B}$ than in the inclusive case; see Fig.~4.  \s

To simulate detector effects, such as acceptance, muon momentum smearing, and
$b$--jet tagging,  we take the example of the CMS detector. Using the CMSJET 
package \cite{CMSJET}, in which muon momentum smearing has been parametrized 
from a full simulation of the CMS  tracker layout as described in 
Ref.~\cite{Tracker},  a mass resolution of about 1\% on the dimuons from the
Higgs decays is obtained.  The events are assumed to be triggered by a double
muon trigger, with a 7 GeV threshold and within an acceptance of $|\eta| <
2.1$, leading to an efficiency of 97\% per muon \cite{Trigger}. For $b$-jet
tagging, we use the b-tagging efficiency obtained by the technique described in
Ref.~\cite{Belluci}. This technique, developed for tagging soft  jets of 20--30
GeV, does not require the $b$-jet reconstruction in the calorimeter, but 
exploits the tracker information [only searching for vertex and tracks  with
significant impact parameter]. With a full detector simulation, such a method
gives 40\% efficiency per $gg \to b\bar{b}\,+\,$Higgs event  for a Higgs boson
mass of 150 GeV.   \s

The results of the simulation for an integrated luminosity of 100 fb$^{-1}$ are
shown in Fig.~6. As expected, the signal invariant mass distributions become
broader, even with the good CMS momentum resolution. This is shown in the plots
of Fig.~6, where the number of $\mu^+\mu^- b\bar{b}$ events in bins of 0.25 GeV
are shown as a function of the mass of the dimuon system. The left-hand side
shows the signals with and without the resolution smearing as obtained  in the
Monte-Carlo analysis, while the  figures in the right-hand side show also the
backgrounds, including the  detector effects. \s

For point P1, the signal cross section for the heavier  CP-even $H$ boson is
significantly smaller than the signals from the lighter CP-even $h$ and
pseudoscalar $A$ bosons; the latter particles are too  too close in mass to be
resolved, and only one single peak for $h/A$ is  clearly visible. To resolve
also the peak for the $H$ boson, the   integrated luminosity should be
increased  by a factor of 3 to 4. In the case of point P2, it would be possible
to see also  the  second peak, corresponding to the $H$ boson signal with a
luminosity  of 100 fb$^{-1}$, but again the $h$ and $A$ peaks cannot be
resolved.  In the case of point P3, all three $h,A$ and $H$ bosons have
comparable signal rates, and the mass differences are large enough for us to
hope to be able to isolate the three different peaks, although with some
difficulty. \s

Note that, if the $\tau^+ \tau^-$ final-state decays of the Higgs bosons 
had been used, with the expected resolution on $\tau$ leptons, we would have
seen only one broad resonance and could not have resolved even two signal
peaks in all three scenarios.

\begin{figure}[!ht]
\begin{center}
\vspace*{-1.1cm}
\centerline{\epsfig{file=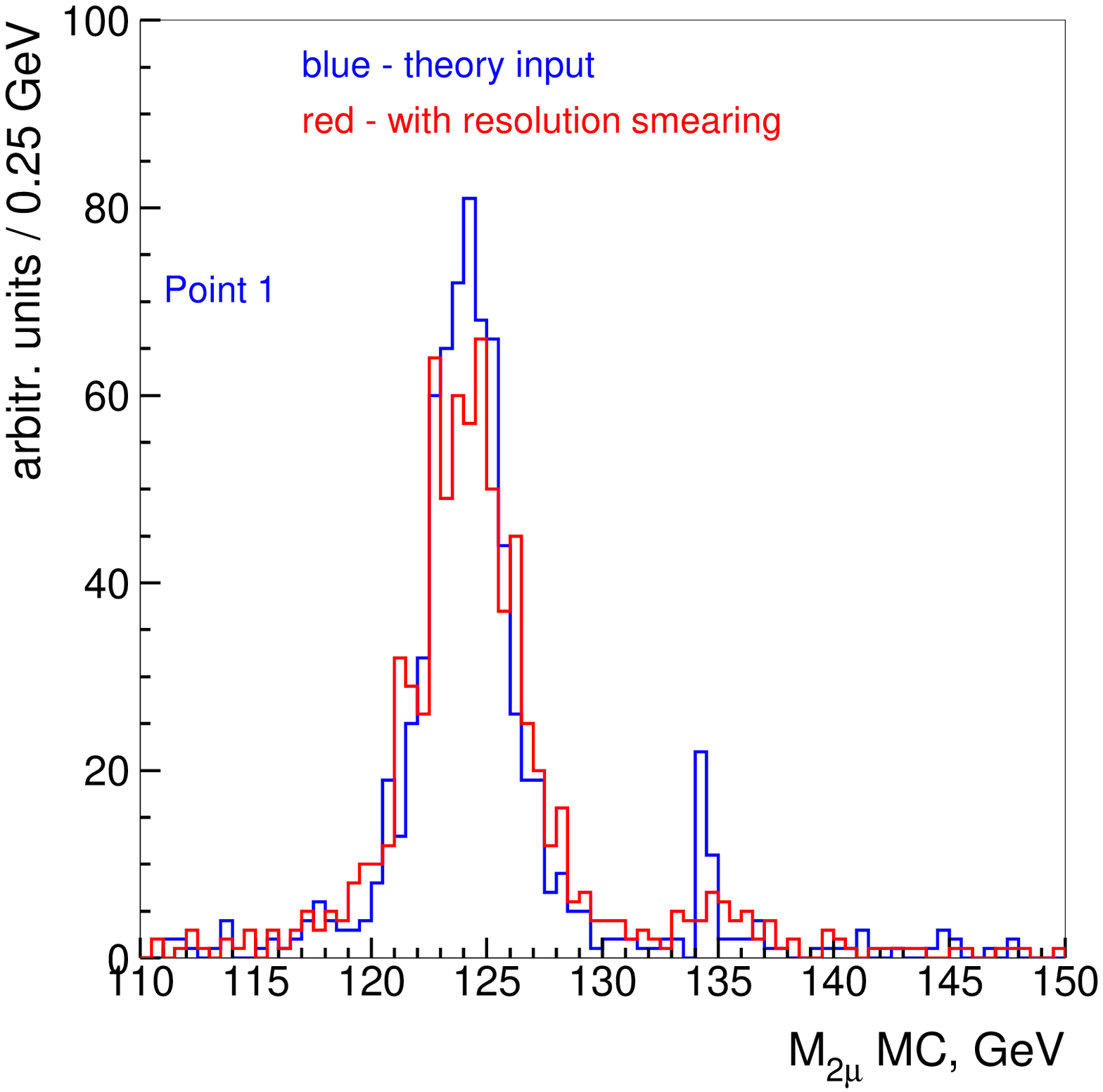,width=7cm}
\epsfig{file=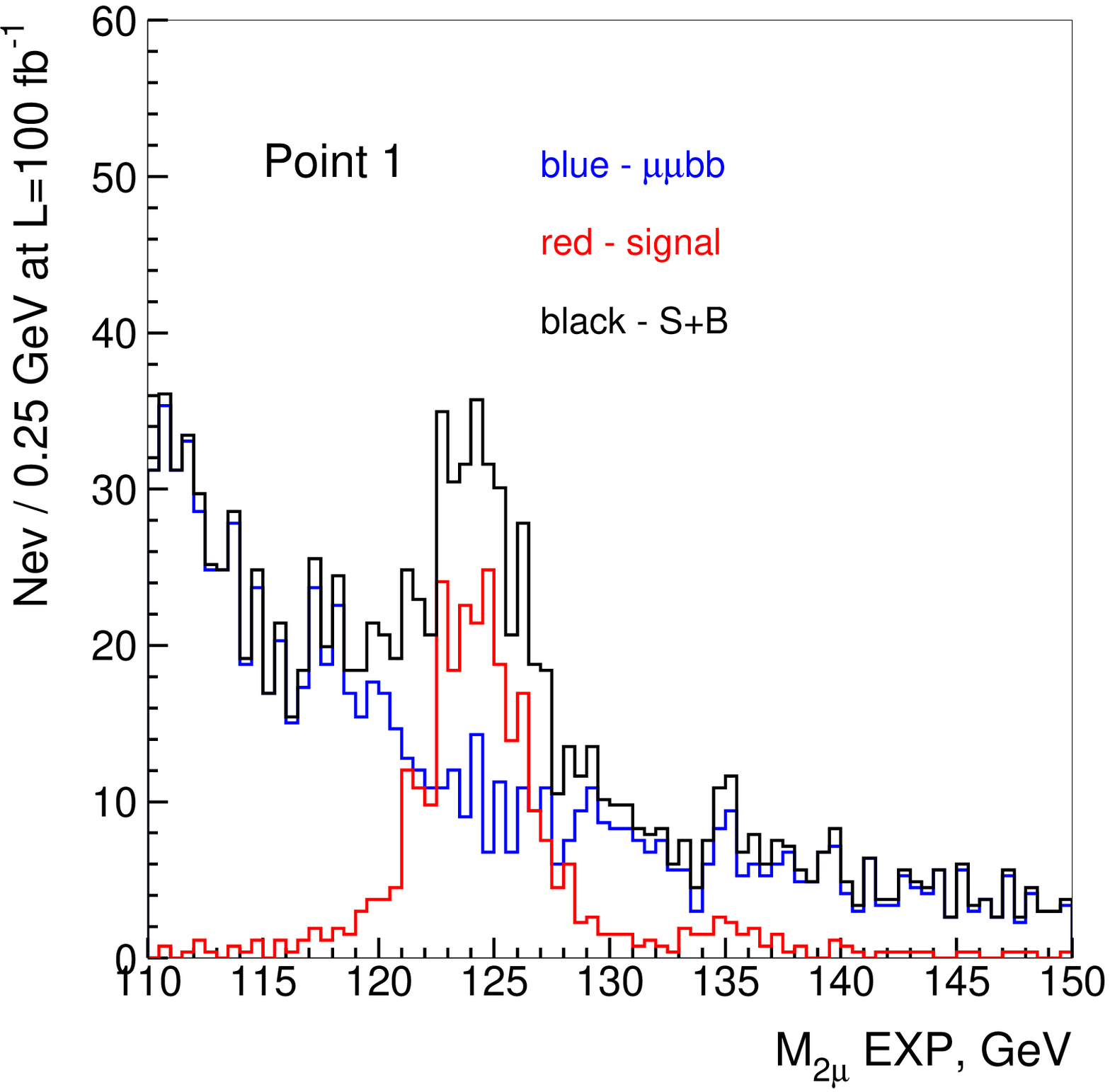,width=7cm}}
\centerline{\epsfig{file=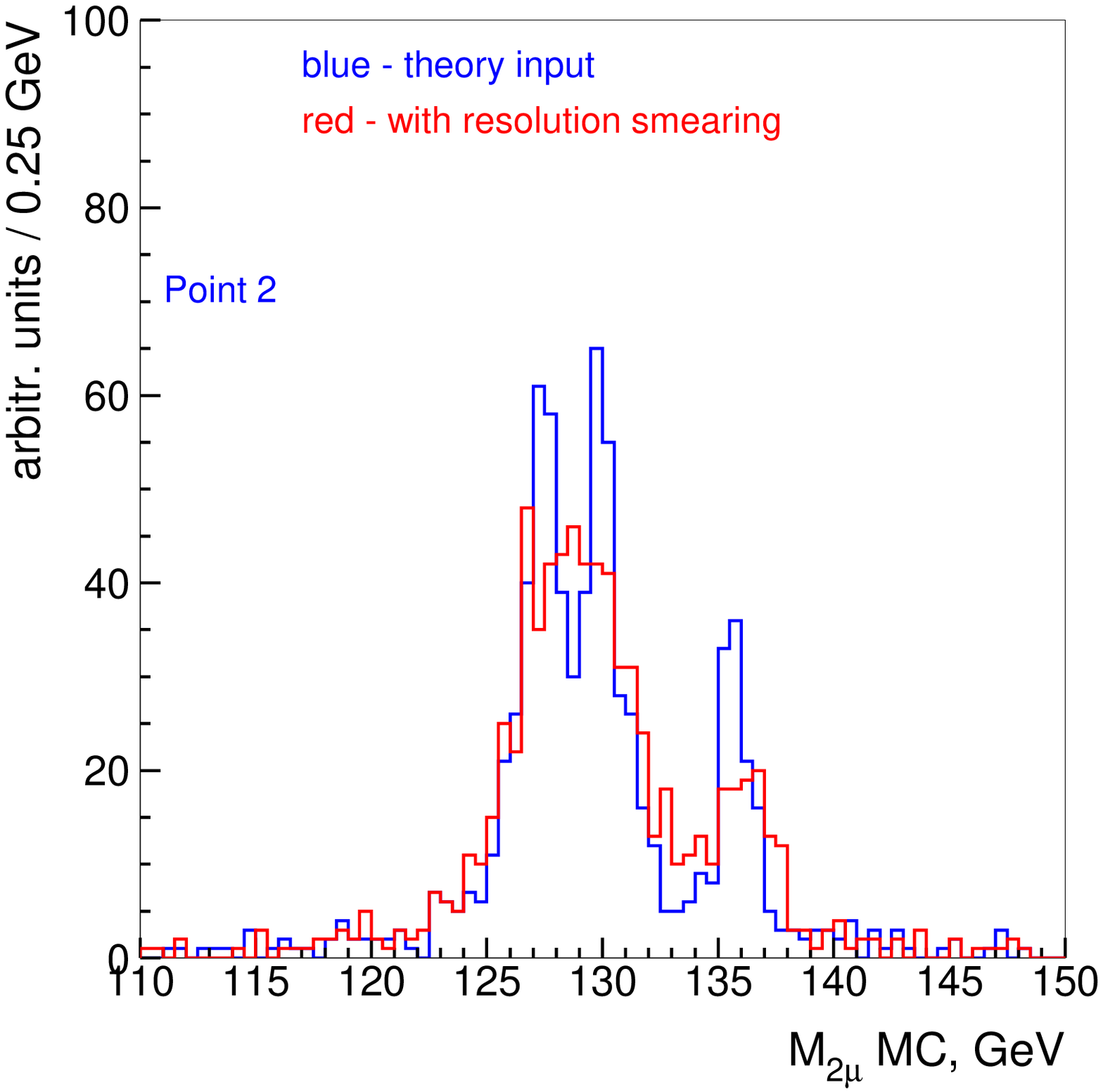,width=7cm}
\epsfig{file=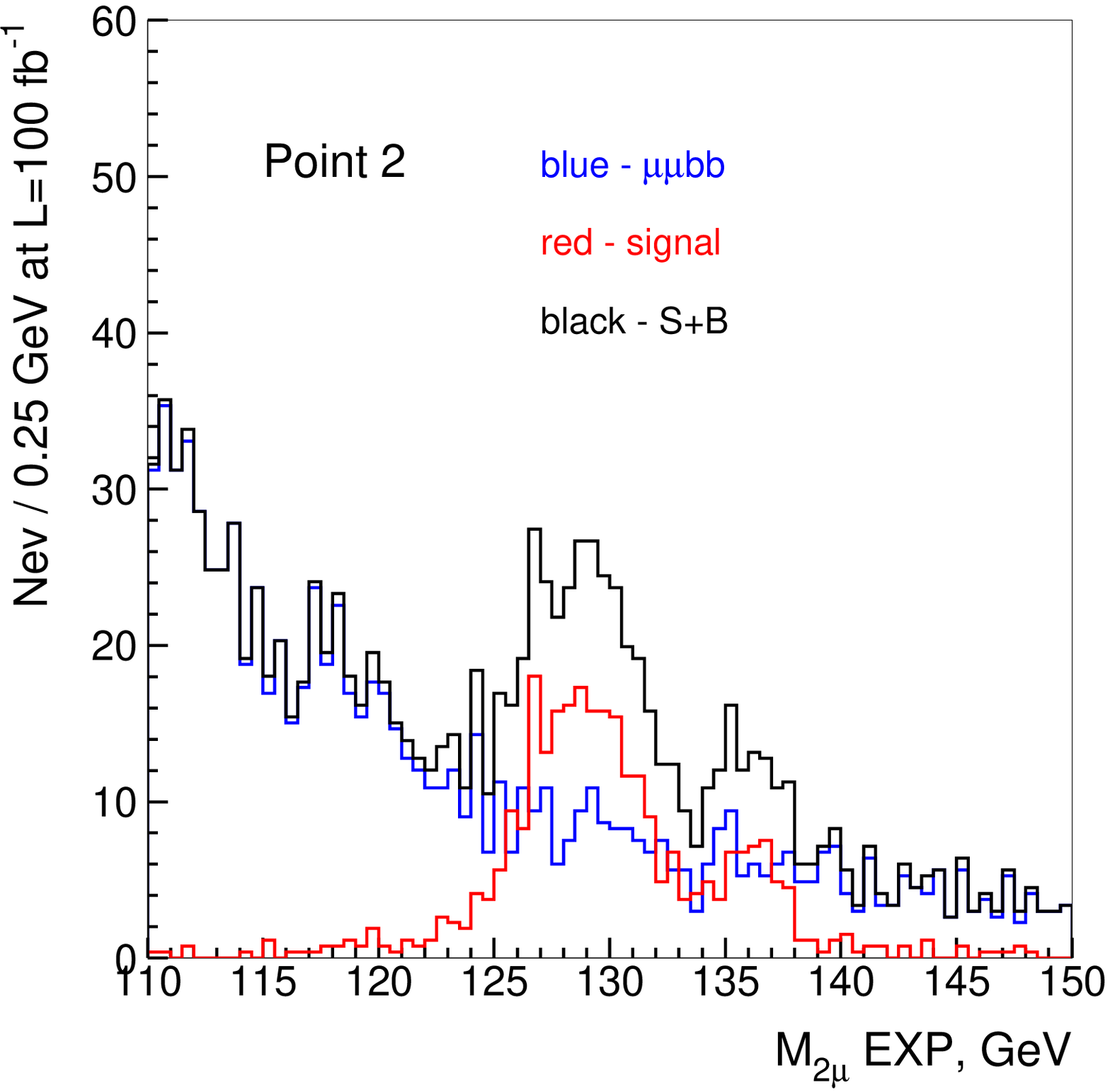,width=7cm}}
\centerline{\epsfig{file=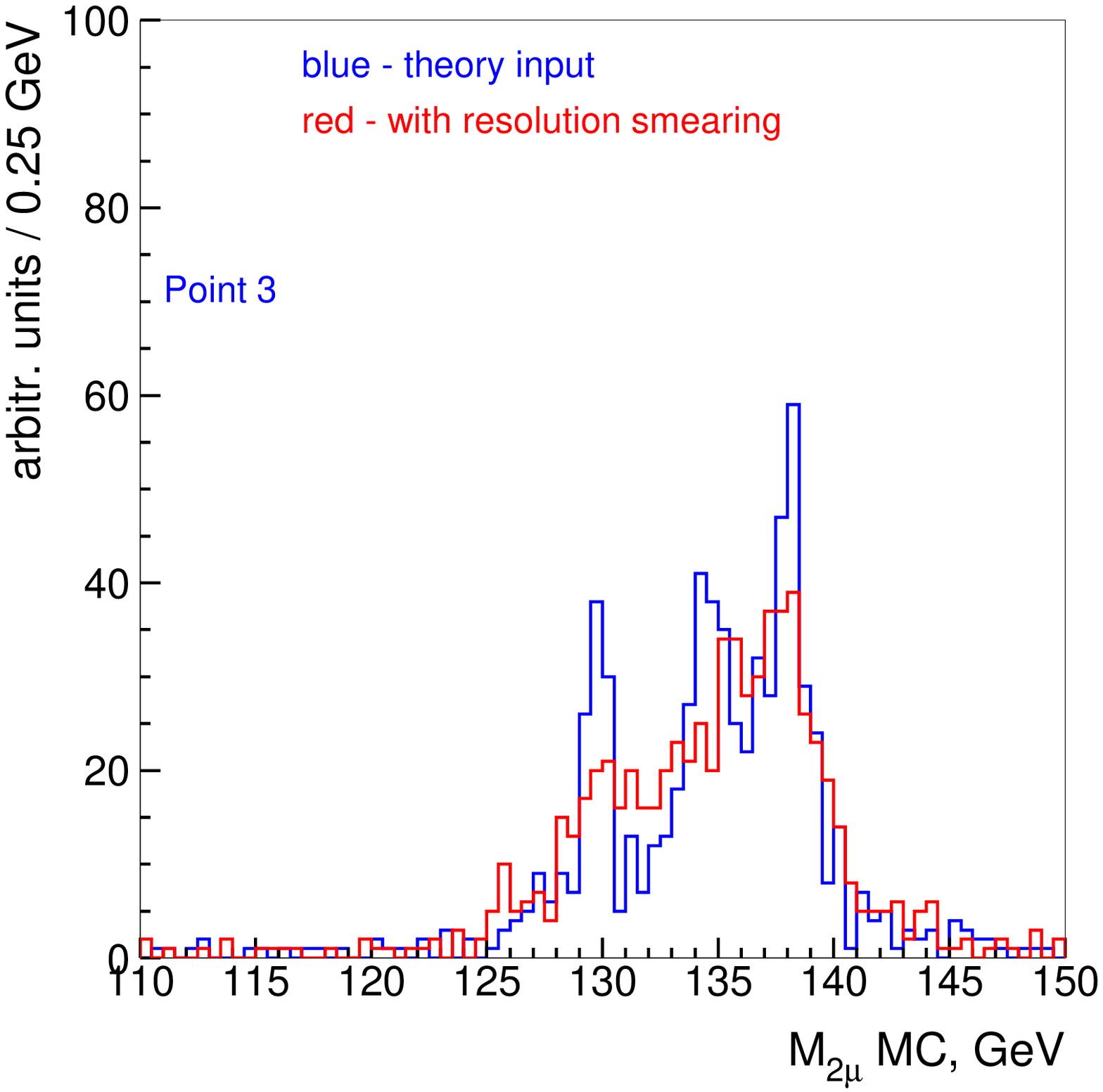,width=7cm}
\epsfig{file=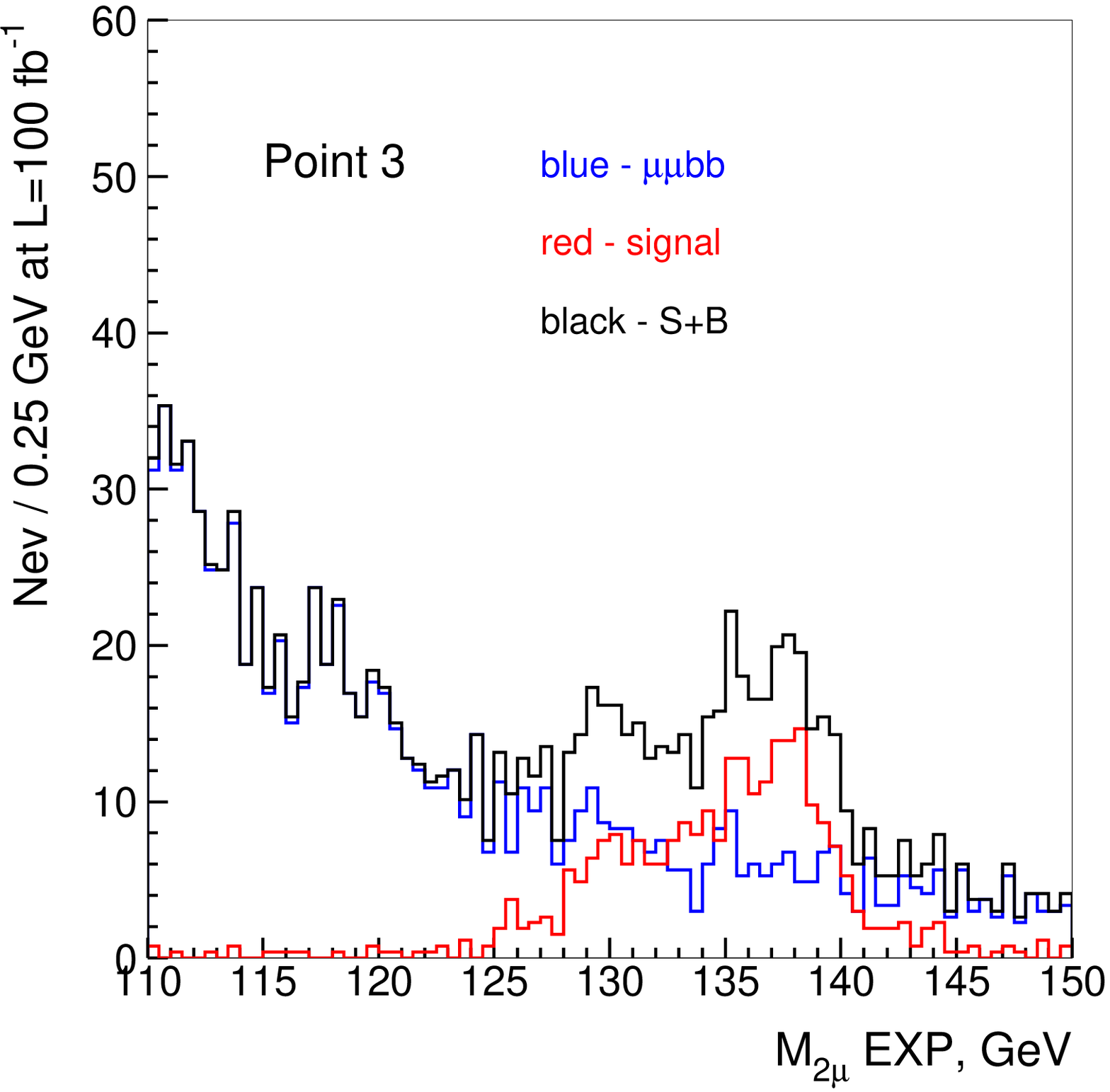,width=7cm}}
\caption{\it $\mu^+ \mu^-$ pair invariant mass distributions
for the signal before and after detector resolution smearing (left)
and for the signal and the background (right) for P1, P2 and P3
parameter points.}
\end{center}
\vspace*{-1.1cm}
\end{figure}
\newpage
 
\subsection*{5. Conclusions}

We have shown that in the intense-coupling regime, i.e. when the $h,H$ and $A$
MSSM bosons have masses too close to the critical point $M_h^{\rm max}$ and
when the value of $\tb$ is large, the detection of the individual Higgs boson
peaks is very challenging at the LHC. It is only in the associated Higgs
production  mechanism with $b\bar{b}$ pairs, with at least one tagged $b$-jet,
and with Higgs particles decaying  into the clean muon-pair final states, that
there is a  chance of observing the three signals and resolve between 
them\footnote{Very recently, it has been argued that in central diffractive 
Higgs production, $pp \to p+\Phi+p$, the cross section for $A$ production 
is very small, and these processes could allow to discriminate between $h$ and 
$H$ production  since, with forward proton taggers, a very small mass
resolution can be obtained \cite{Diffractive}.}. This would be possible only if
the Higgs  mass differences are larger than 3--5 GeV. \s

In the present note, we only concentrated on the fully exclusive $b\bar{b}+
\mu^+\mu^-$ signature, requiring both $b$-jets to be observed, and included
only the irreducible 4-fermion background, which is expected to be the dominant
one. In a more complete study,  one should eventually consider the case where
only one single $b$-jet is tagged, which should increase the cross section
signal \cite{Scott}, and take into  account also the reducible backgrounds from
$pp \to Z^*/\gamma^* \to \mu^+\mu^-$ with mistagged jets [which is expected  to
be large in this case] as well as other backgrounds. Such a study is beyond the 
scope of this note and will appear elsewhere \cite{elsewhere}. \bigskip

\nn {\bf Acknowledgments:} We thank John Campbell, Michael Kramer, Tilman
Plehn, Laura Reina, Michael Spira and Scott Willenbrock for very lively
discussions on the $b\bar{b}$\,+\,Higgs process, during and after the Les
Houches Workshop.   Discussions with Daniel Denegri and Elzbieta Richter-Was
are also acknowledged. EB is partly supported by the INTAS 00-0679, CERN-INTAS
99--377 and Universities of Russia UR.02.03.002 grants.

\end{document}